\documentclass[12pt,letter]{article}
\usepackage{natbib}
\usepackage{amsmath}
\usepackage[showonlyrefs]{mathtools}
\usepackage{amssymb}
\usepackage{graphicx,psfrag,epsf}
\usepackage{subfig}
\RequirePackage{amsfonts}
\usepackage{times}
\usepackage{enumerate}
\usepackage{multirow} 
\usepackage{url} 
\usepackage{version}
\usepackage{setspace}
\usepackage{nicefrac}
\usepackage{algorithm}
\usepackage{algpseudocode}
\usepackage{varwidth}
\usepackage{xspace}
\usepackage{bm}
\usepackage{tikz}
\usetikzlibrary{shapes,snakes}
\usepackage{pgfplots}
\newcommand{\blind}{0}

\newcommand{\tit}{Adaptive sequential Monte Carlo for multiple changepoint analysis}

\definecolor{light-gray}{gray}{0.65}

\newcommand{\I}{\mathbb{I}}

\newcommand{\diff}{\mathrm{d}}
\newcommand{\ns}{m}

\newcommand{\US}{SMCcp\xspace}
\newcommand{\DM}{SMCs\xspace}
\newcommand{\NW}{SMCpdp\xspace}
\newcommand{\AJ}{\mathrm{\DM}}
\newcommand{\argmax}{\arg \! \max}

\newcommand{\ttau}{\tilde{\tau}}

\newcommand{\kk}{\tilde{k}}
\newcommand{\kki}{\tilde{k}}
\newcommand{\ki}{k}
\newcommand{\qp}{q}
\newcommand{\ttheta}{\tilde{\theta}}
\newcommand{\tlambda}{\tilde{\lambda}}
\newcommand{\tpi}{\tilde{\pi}}
\newcommand{\tzero}{0}
\newcommand{\istar}{i^\ast}
\newcommand{\istartwo}{i^{\ast\ast}}
\newcommand{\tokornottok}[1]{{\boldsymbol{\tau}}_{#1}}
\newcommand{\simplifiedtheta}[1]{\boldsymbol{\theta}_{#1}}
\newcommand{\btokornottok}[1]{{\boldsymbol{\ttau}}_{#1}}
\newcommand{\bsimplifiedtheta}[1]{\boldsymbol{\ttheta}_{#1}}
\newcommand{\simplifiedlambda}[1]{\boldsymbol{\lambda}_{#1}}
\newcommand{\bsimplifiedlambda}[1]{\boldsymbol{\tilde{\lambda}}_{#1}}
\begin{document}
\def\spacingset#1{\renewcommand{\baselinestretch}%
{#1}\small\normalsize} \spacingset{1}
\if0\blind
{
  \title{\bf \tit}
  \author{Melissa J. M. \textsc{Turcotte}\thanks{The first author gratefully acknowledges support from the EPSRC and the Institute for Security Science and Technology, Imperial College London.}\hspace{.2cm}\\
    Advanced Computing Solutions and the Center for Nonlinear Studies, \\Los Alamos National Laboratory\\
    and\\
    Nicholas A. \textsc{Heard}\\
    Department of Mathematics, Imperial College London}
  \maketitle
}\fi


\if1\blind
{
  \bigskip
  \bigskip
  \bigskip
  \begin{center}
    {\LARGE\bf \tit}
\end{center}
  \medskip
} \fi

\bigskip
\begin{abstract}
Process monitoring and control requires detection of structural
changes in a data stream in real time. This article introduces an
efficient sequential Monte Carlo algorithm designed for
learning unknown changepoints in continuous time. The method is
intuitively simple: new changepoints for the latest window of data are
proposed by conditioning only on data observed since the most recent
estimated changepoint, as these carry most of the information about
the state of the process prior to the update. The proposed method
shows improved performance over the current state of
the art.

Another advantage of the proposed algorithm is that it can be made
adaptive, varying the number of particles according to the apparent local
complexity of the target changepoint probability distribution. This saves valuable
computing time when changes in the changepoint distribution
are negligible, and enables re-balancing of the importance weights of
existing particles when a significant change in the target
distribution is encountered.

The plain and adaptive versions of the method are illustrated using
the canonical continuous time changepoint problem of inferring the
intensity of an inhomogeneous Poisson process. Performance
is demonstrated using both conjugate and non-conjugate Bayesian models
for the intensity.
\end{abstract}

\noindent%

\spacingset{1.45} 

\section{Introduction}\label{sec:intro}

Let $\{y(t):t\geq \tzero\}$ be a continuous time stochastic process on
$\mathbb{R}^+$, where the law of $y(t)$ is governed by a second
underlying stochastic process $\theta(t)\in\Theta$. A changepoint
model for $y$ assumes an unknown number of changepoints
$\tau_1<\tau_2<\ldots$ that partition $\mathbb{R}^+$ into disjoint,
homogeneous segments $[\tau_i,\tau_{i+1})$ such that 
  $\forall t \in [\tau_i,\tau_{i+1}), \theta(t)=\theta_i$ for some
    constant parameter value $\theta_i$. Changepoint detection is
    concerned with finding the number and location of changepoints in
    $\theta(t)$ over a finite observation period $[0,T]$, and in a
    sequential setting, detecting each change soon after it has
    occurred.

The motivation behind the sequential Monte Carlo (SMC) algorithm
proposed here is detecting changepoints in continuous time,
previously considered by \cite{whiteley11} under the name
\textit{piecewise deterministic processes}. The process 
data $y(\cdot)$ are assumed to arrive as a continuous stream, while inferences
about $\theta(\cdot)$ 
are made at a discrete sequence of observation times
$\tzero<t_1<t_2<\ldots$. At each observation time $t_n$, $n\geq 1$,
the posterior distribution for changepoints in the interval
$[\tzero,t_n]$ is sought in order to make such inferences.

The SMC algorithm of \cite{whiteley11} is a direct application of the
SMC samplers methodology of \cite{ajay06}, which is a more general SMC
technique for sampling sequentially from \textit{any} sequence of
target distributions defined on a common space. The generality of the
SMC samplers method is achieved by augmenting the target distributions
to ever increasing dimensions in order to avoid the need to integrate
over a general transition kernel; whilst that provides the basis for a
very general class of samplers, applicable in a broad variety of
contexts, the aim of this article is to propose a more bespoke SMC
algorithm designed specifically for changepoint analysis.

In the wider literature of SMC for changepoint detection, but within
the context of discrete time changepoint analysis, \cite{fearnhead07}
make sequential inference on data where filtering recursions are used
to sample exactly from the distribution of the most recent
changepoint, and consequently the joint distribution of all
changepoints; the computational cost of exact simulation increases
linearly with time and so an approximation using particle filtering is
proposed. \cite{chopin07} and \cite{fearnhead03} approach discrete
changepoint detection in time series by reformulating the changepoint
problem as a hidden Markov model and use particle filters to propagate
forward the distribution of the time since the most recent
changepoint. Carrying forward inferences about the most recent
changepoint is also an important element of the proposed method.

The next section outlines the Bayesian changepoint model. Section
\ref{sec:SMCcp} proposes an efficient new SMC algorithm for
changepoint problems; Section \ref{sec:adaptive_smc} demonstrates how
this algorithm can easily be made adaptive, automatically varying the
number of particles according to the complexity of the target, which
can potentially be valuable when performing inference on a number of
sequences of target distributions in parallel.

\section{Bayesian changepoint model}

In all of the examples in this article, the arrivals of changepoints in $\theta(t)$ will be assumed to be a homogeneous Poisson process with rate $\nu\in \mathbb{R}^{+}$. Following the notation of \cite{whiteley11}, let $k_n$ be the number of
changepoints over $[\tzero,t_n]$, and define
$\tau_{0}=\tzero$ and $\tau_{ k_n+1}=t_n$. If $k_n=0$ then let
$\tau_{1:k_n}=0$, otherwise let $\tau_{1:k_n}=(\tau_{ 1},\ldots,\tau_{
  k_n})$ denote the ordered locations of these changepoints, so that
$\tau_{ 1:k_n}\in \mathbb{T}_{n,k_n}$ where
$\mathbb{T}_{n,k_n}=\{\tau_{
  1:k_n}:\tzero<\tau_{1}<\ldots<\tau_{k_n}<t_n\}.$
 Defining $\boldsymbol{\tau}_n=(k_n,\tau_{1:k_n})$ the prior density on $[\tzero,t_n]$ implied by the Poisson process is
\begin{equation*}
p(\tokornottok{n})=\nu^{k_n}\exp(-\nu t_n)\I_{\mathbb{T}_{n,k_n}}(\tau_{1},\ldots,\tau_{k_n}).
\label{eq:priorchangepoints}
\end{equation*}

Let  $\simplifiedtheta{n}=\theta_{0:k_n}=(\theta_{0},\ldots,\theta_{ k_n})\in\Theta^{k_n+1}$ be the vector of corresponding parameters for the $k_n+1$ segments of the partition of $[\tzero,t_n]$ created by the changepoints $\tau_{1:k_n}$. Denote by $y([\tzero,t_n])$ the sample path $y(t)$ over $[\tzero,t_n]$. Assuming a likelihood function
 $f\left(y([\tzero,t_n])|\tokornottok{n}, \simplifiedtheta{n} \right)$ for the sample path which is known pointwise, and a prior distribution for the
model parameters $p\left(\simplifiedtheta{n}|\tokornottok{n}\right)$, the joint density of the changepoints, the
parameters and the sample path is immediately available as
\begin{equation}
\gamma_{[\tzero,t_n]}(\tokornottok{n},\simplifiedtheta{n}, y([\tzero,t_n]))= f\left(y([\tzero,t_n])|\tokornottok{n},  \simplifiedtheta{n} \right)p\left(\simplifiedtheta{n}|\tokornottok{n}\right)p\left(\tokornottok{n}\right).
\label{eq:posterior_nonconjugate_approx}
\end{equation}
The posterior density for the changepoints and parameters is then known up to proportionality by
\begin{equation}\pi_{[\tzero,t_n]}(\tokornottok{n},\simplifiedtheta{n}|y([\tzero,t_n]))\propto \gamma_{[\tzero,t_n]}(\tokornottok{n},\simplifiedtheta{n}, y([\tzero,t_n])).
  \label{eq:posterior_nonconjugate}
\end{equation}
Generally interest centers on estimating the posterior density for the changepoints, whereas the parameters $\simplifiedtheta{n}$ are regarded as nuisance parameters. If the conditional density of the parameters given the changepoints is the conjugate prior to the likelihood model, then the parameters $\simplifiedtheta{n}$ can be marginalised out. The posterior density for the changepoints then has the form
\begin{equation}
\pi_{[\tzero,t_n]}(\tokornottok{n}|y([\tzero,t_n]))\propto\gamma_{[\tzero,t_n]}(\tokornottok{n}, y([\tzero,t_n])),
\label{eq:posterior}
\end{equation}
where the joint density in \eqref{eq:posterior} is a
marginal of \eqref{eq:posterior_nonconjugate_approx} with respect to
the parameter vector $\simplifiedtheta{n}$. The
sequence of these posterior densities has support over nested
transdimensional spaces $E_n \subset E_{n+1}$ where
\begin{equation*}
E_n=\bigcup_{k_{ n}=0}^{\infty}\{k_{ n}\}\times\mathbb{T}_{ n, k_{ n}}.
\label{eq:space}
\end{equation*}

\section{Sequential Monte Carlo algorithm for changepoint distributions}\label{sec:SMCcp}

To make a computationally fast SMC algorithm for changepoint
problems, the proposal distributions for the sequence of target
distributions $\pi_{[\tzero,t_n]}$, for $n=1{,}2{,}\ldots$, will
sample the changepoints for each update interval $(t_{n-1},t_n]$
  without reference to the sampled changepoints from the previous
  intervals, but will instead condition on the data observed since an
  earlier time $t^\ast_{n-1}\leq t_{n-1}$, where $t^\ast_{n-1}$
  should be interpreted as some convenient estimate of $\tau_{k_{n-1}}$, the
  most recent changepoint in $[\tzero,t_{n-1}]$
  \citep[\textit{cf.}][]{chopin07, fearnhead03}. Here,
  $t^\ast_{n-1}$ will be the posterior mean of $\tau_{k_{n-1}}$
  estimated from the weighted SMC sample at $t_{n-1}$. Other
  choices are possible: the straightforward choice of
  $t^\ast_{n-1}=t_{n-1}$ was also tested, but led to reduced
  performance in all examples.

The motivation behind the proposed algorithm is as follows: If new
changepoints within $(t_{n-1},t_n]$ are to be sampled and appended to
the existing changepoint vectors for each particle, then for
constructing a good proposal distribution it might be sufficient to retain
only those data that have been observed since the last changepoint in
$[\tzero,t_n]$, as these data provide all of the available information
on the current state of $\theta(t)$ as the process enters
$(t_{n-1},t_n]$. The diversity of the particles in terms of their
  earlier changepoints in $[\tzero,t_{n-1}]$ has no future bearing.

Let $\kk_{n}\leq k_n$ be a random variable for the number of
changepoints in the \textit{update interval} $(t_{n-1},t_n]$ and let  $\boldsymbol{\ttau}_n=(\kk_n,\ttau_{1:\kk_n})$. Define
\begin{equation}\pi_{ (t_{n-1},t_n]}(\btokornottok{n}|y((t^\ast_{n-1},t_n]))]\propto\gamma_{(t_{n-1},t_n]}(\btokornottok{n}, y((t^\ast_{n-1},t_n])),
\label{eq:posterior_interval}
\end{equation}
as the local posterior distribution of changepoints on the subspace
$$
\tilde{E}_n=\bigcup_{\kk_{ n}=0}^{\infty}\{\kk_{ n}\}\times \{\ttau_{ 1:\kk_n}:t_{n-1}<\ttau_{1}<\ldots<\ttau_{\kk_n}<t_n\}
$$ when conditioning only on the data observed in the extended interval $(t^\ast_{n-1},t_n]$. 

  Reversible jump Markov chain Monte
  Carlo (RJMCMC) \citep{green05} sampling from the densities
  \eqref{eq:posterior_interval} will provide the proposals
  in the SMC algorithm. Importantly, convergence of
  RJMCMC on the subintervals $(t_{n-1},t_n]$ will be very fast if the update intervals are small,
  as the prior probability that there would be more than one changepoint is $o(t_n-t_{n-1})$.



Extending a vector of changepoints in the case of a conjugate Bayesian model,
with target density \eqref{eq:posterior}, is the most straightforward
case and will be considered first. Second, the non-conjugate case will
be addressed, with extended target density
\eqref{eq:posterior_nonconjugate}. This latter case is more difficult
for two reasons: firstly, sampling from more highly parameterized
models is generally more cumbersome and inefficient, but secondly and
more importantly, when appending changepoint vectors from the update
interval there is a spare \textit{intercept} parameter that needs to
be properly handled.

\subsection{Conjugate models}\label{sec:conjugate_models}
Algorithm \ref{PFSMC} presents the most straightforward form of the
SMC algorithm being proposed, when assuming a conjugate Bayesian model
for $y(t)$ within each changepoint segment.
\begin{algorithm}[h]\caption{SMC algorithm for changepoint detection}{\label{PFSMC}}
\begin{algorithmic}[1]

\State Sample $\{\tokornottok{1}^{(i)}\}_{i=1}^N\sim \pi_{ [\tzero,t_1]}(\tokornottok{1}|y([\tzero,t_1]))$ via RJMCMC and set $w^{(i)}_{1}=1$ for $i=1,\ldots,N$ 

\State Set ${\it n=2}$

\State Calculate $t^\ast_{n-1}=\sum_{i=1}^N w_{n-1}^{(i)} \tau_{k_{n-1}}^{(i)}/\sum_{i=1}^N w_{n-1}^{(i)}$\label{gotoline}

\State Sample $\{\btokornottok{n}^{(i)}\}_{i=1}^{N}\sim \pi_{ (t_{n-1},t_n]}(\btokornottok{n}|y((t^\ast_{n-1},t_n]))$ via RJMCMC 

\State Draw a random permutation $\sigma$ uniformly from $S_N$, the symmetric group on $N$ symbols\label{step:permute}

\State Combine the particles, $\{\tokornottok{n}^{(i)}=(k_{n-1}^{(i)}+\kk_{n}^{(i)},\tau_{1}^{(i)},\ldots,\tau_{\ki_{n-1}}^{(i)},\ttau_{1}^{\sigma(i)},\ldots,\ttau_{\kki_{n}}^{\sigma(i)})\}_{i=1}^{N}$ \label{step:combine}

\State Update the importance weights $\{w_n^{(i)}\}_{i=1}^N$ according to \eqref{eq:weight}

\State Calculate the effective sample size $\mbox{ESS}=\sum_{i=1}^N {w_{n-1}^{(i)}}^2 /(\sum_{i=1}^N w_{n-1}^{(i)})^2$ \label{step:ess}

\If {ESS $<\mbox{ESS}_{T}=N/3$}

\State Resample $\{\tokornottok{n}^{(i)}, w^{(i)}_{n}\}_{i=1}^N$ according to the weights to obtain unweighted particles $\{\tokornottok{n}^{(i)}, 1\}_{i=1}^{N}$\label{step:ESSresample}
\State Optionally move each particle according to a
$\pi_{[\tzero,t_n]}$ invariant kernel and retain the same importance weights\label{step:move}
\EndIf
\State ${\it n\leftarrow n+1}$

\State {\bf goto} \ref{gotoline}
\end{algorithmic}
\end{algorithm}

At time $t_{n-1}$, the algorithm assumes a set of $N$ importance
weighted particles
$\{\tokornottok{n-1}^{(i)},w_{n-1}^{(i)}\}_{i=1}^{N}$ that
approximate $\pi_{[\tzero,t_{n-1}]}(\tokornottok{n-1}|y([0,t_n]))$. Let
\begin{equation*}
t^\ast_{n-1}=\sum_{i=1}^N w_{n-1}^{(i)} \tau_{k_{n-1}}^{(i)}\bigg/\sum_{i=1}^N w_{n-1}^{(i)}
\end{equation*}
be the Monte Carlo estimate of the posterior mean of $\tau_{k_{n-1}}$.
Then at time $t_{n}$, $N$ new
sub-particles are sampled independently from $\pi_{
  (t_{n-1},t_n]}(\btokornottok{n}|y((t^\ast_{n-1},t_n]))$ 
    \eqref{eq:posterior_interval} via RJMCMC.
Then setting $\tokornottok{n}=(\tokornottok{n-1},\btokornottok{n})$, the implied importance distribution on $[\tzero,t_n]$ from combining the two sets of particles is known only up to proportionality through 
\begin{align}
\qp_{ [\tzero,t_n]}(\tokornottok{n})&=\gamma_{[\tzero,t_{1}]}(\tokornottok{1},y([\tzero,t_1]))\prod_{j=2}^{n}\gamma_{(t_{j-1},t_{j}]}(\btokornottok{j},y((t^\ast_{n-1},t_n])).\nonumber
\label{eq:proposal}
\end{align}

Since RJMCMC is used to sample from $\pi_{
  (t_{n-1},t_n]}(\btokornottok{n}|y((t^\ast_{n-1},t_n]))$ in
    each update, there will be autocorrelation in each batch of
    samples. To negate this, before joining the particles together
    step \ref{step:permute} of the algorithm permutes the labels of
    the sample from the new interval, to break the autocorrelation of
    the combined particles.

The importance weights are then updated to account for the discrepancy between the importance distribution and the changepoint posterior distribution \eqref{eq:posterior}. For $n>1$ the weight for the $i$th particle is given by
\begin{equation}
w_{n}(\tokornottok{n}^{(i)})=\displaystyle\frac{\gamma_{{[\tzero,t_n]}}(\tokornottok{n}^{(i)},y([\tzero,t_n]))}{\qp_{[\tzero,t_n]}(\displaystyle\tokornottok{n}^{(i)})}=w_{n-1}(\tokornottok{n-1}^{(i)})\overline{w}_{n}(\tokornottok{n}^{(i)}),
\label{eq:weight}
\end{equation}
where the incremental weight
\begin{equation}
\overline{w}_{n}(\tokornottok{n}^{(i)})=\frac{\gamma_{{[\tzero,t_n]}}(\tokornottok{n}^{(i)},y([\tzero,t_n]))}{\gamma_{[\tzero,t_{n-1}]}(\tokornottok{n-1}^{(i)},y([\tzero,t_{n-1}]))\gamma_{(t_{n-1},t_n]}(\btokornottok{n}^{(i)}, y((t^\ast_{n-1},t_n]))},
\label{eq:incrementalweight}
\end{equation}
is computationally simple to calculate; for example, in the special
case of $t^\ast_{n-1}=t_{n-1}$ this is analogous to calculating
the probability of accepting a reversible jump death move for a
changepoint at $t_{n-1}$.

Note that the entire SMC scheme is equivalent to sequential importance sampling on a sequence of distributions defined on a common space with a transition kernel
\begin{equation*}
K_n(\tokornottok{n}|\tokornottok{n-1}^{'})\propto\delta_{\tokornottok{n-1}^{'}}(\tokornottok{n-1})\gamma_{(t_{n-1},t_{n}]}(\btokornottok{n},y((t^\ast_{n-1},t_n])).
\label{eq:kernel}
\end{equation*}
 
When the \textit{effective sample size} (ESS) \citep{liu06} of the particles drops
below a threshold, commonly taken to be $N/3$, the systematic resampling
approach \citep{kitagawa96} is used (Algorithm \ref{PFSMC}, Step
\ref{step:ess} to \ref{step:ESSresample}). Additionally, a sweep of RJMCMC moves is applied to
the particle set after resampling (Algorithm \ref{PFSMC}, Step \ref{step:move}), as adopted in both \cite{ajay06}
and \cite{whiteley11}.

\subsection{Non-conjugate models}\label{sec:non-conjugate}
The SMC algorithm for non-conjugate models follows the steps of Algorithm \ref{PFSMC}, with the exception that the marginal posteriors $\pi_{[\tzero,t_n]}(\tokornottok{n}|y([\tzero,t_n]))$ from \eqref{eq:posterior} are unavailable. Changepoints are sampled at each update via RJMCMC, but from the joint distribution of changepoints and parameters $\pi_{ (t_{n-1},t_n]}(\btokornottok{n},\bsimplifiedtheta{n}|y((t^\ast_{n-1},t_n]))$.

Assume at $t_{n-1}$ a weighted sample of changepoints and parameters $\{\tokornottok{n-1}^{(i)},\simplifiedtheta{n-1}^{(i)},w_{n-1}^{(i)}\}_{i=1}^{N}$ has been obtained from $\pi_{[\tzero,t_{n-1}]}(\tokornottok{n-1},\simplifiedtheta{n-1}|y([0,t_{n-1}]))$, and subsequently at time $t_n$ a sample $\{\btokornottok{n}^{(i)},\bsimplifiedtheta{n}^{(i)}\}_{i=1}^{N}$ is drawn from $\pi_{(t_{n-1},t_{n}]}(\btokornottok{n},\bsimplifiedtheta{n}|y((t^\ast_{n-1},t_n]))$. Combining the particles from these two samples is now less straightforward, as there is an extra, redundant parameter for $\theta(t)$ for the segment $(\tau_{k_{n-1}},\ttau_1]$. The implied proposal distribution would be over-parameterized, so the parameter pair $(\theta_{k_{n-1}},\ttheta_{0})$ needs to be combined to form a single parameter $\theta^{*}_n$. Let $s_1(\theta_{k_{n-1}},\ttheta_{0})$ be a suitably chosen function to combine the two model parameters into a single value $\theta^{*}_n$.

As the marginal distribution of $\theta^{*}_n$ implied by the proposal density and $s_1$ is unlikely to have an analytic solution, a joint change of variables is required. Let $s_2(\theta_{k_{n-1}},\ttheta_{0})$ be a second transformation such that the pair
$$
(\theta^{*}_n,u_{ n-1})=s(\theta_{k_{n-1}},\ttheta_{0})=(s_{1}(\theta_{k_{n-1}},\ttheta_{0}),s_{2}(\theta_{k_{n-1}},\ttheta_{0}))
$$
comprise a one to one mapping $(\theta_{k_{n-1}},\ttheta_{0})\mapsto(\theta^{*}_n,u_{ n-1})$, and let $|\mbox{J}_s|$ be the determinant of the Jacobian of $s$. 

The implied proposal density
, following the change of variable $s$, known up to a constant of proportionality satisfies
\begin{align*}
\qp_{ [\tzero,t_n]}(\tokornottok{n},\simplifiedtheta{n},u_{1:n-1})=&\qp_{ [\tzero,t_{n-1}]}(\tokornottok{n-1},\simplifiedtheta{n-1},u_{1:{n-2}})
\gamma_{{ (t_{n-1},t_n]}}(\btokornottok{n},\bsimplifiedtheta{n},y((t^\ast_{n-1},t_n]))|\mbox{J}_s|,
\end{align*}
where $\simplifiedtheta{n}=(\theta_0,\ldots,\theta_{k_{n-1}-1},\theta^*_n,\ttheta_1,\ldots,\ttheta_{\kk})$. This proposal density generates suitable parameters to correspond to the combined changepoints, but also the nuisance parameters $u_{ 1:n-1}$. To accommodate these nuisance parameters,
a general solution is to extend the target distribution
so that
\begin{equation}
\pi_{[\tzero,t_n]}(\tokornottok{n},\simplifiedtheta{n},u_{ 1:n-1}|y([0,t_n]))\triangleq\pi_{[\tzero,t_n]}(\tokornottok{n},\simplifiedtheta{n}|y([0,t_n]))\prod_{j=1}^{n-1}\tpi(u_{j}|\tokornottok{j},\simplifiedtheta{j})\label{eq:posterior_extended}
\end{equation}
where $\tpi$ can be \textit{any} density with the correct support for
$u_{ j}$. As the true target \eqref{eq:posterior_nonconjugate} is a
marginal of \eqref{eq:posterior_extended}, standard importance
sampling estimates obtained in the extended space can still be used to
give an approximation for the true target distribution and its
normalizing constant.

The importance weights given in \eqref{eq:weight} for the non-conjugate case with this extended target are 
\begin{align*}
w_{n}(\tokornottok{n}^{(i)},\simplifiedtheta{n}^{(i)},u_{1:n-1}^{(i)})&=\displaystyle\frac{\gamma_{{[\tzero,t_n]}}(\tokornottok{n}^{(i)},\simplifiedtheta{n}^{(i)},u_{1:n-1}^{(i)},y([\tzero,t_n]))}{\qp_{[\tzero,t_n]}(\displaystyle\tokornottok{n}^{(i)},\simplifiedtheta{n}^{(i)},u_{1:n-1}^{(i)})}\\
&=w_{n-1}(\tokornottok{n-1}^{(i)},\simplifiedtheta{n-1}^{(i)},u_{1:n-2}^{(i)})\overline{w}_{n}(\tokornottok{n}^{(i)},\simplifiedtheta{n}^{(i)},u_{n-1}^{(i)}),
\end{align*}
with the incremental weight
\begin{equation*}
\overline{w}_{n}(\tokornottok{n}^{(i)},\simplifiedtheta{n}^{(i)},u_{n-1}^{(i)})=\frac{\gamma_{ {[\tzero,t_n]}}(\tokornottok{n}^{(i)},\simplifiedtheta{n}^{(i)},y([\tzero,t_n]))\tpi(u_{n-1}^{(i)}|\tokornottok{n}^{(i)},\simplifiedtheta{n}^{(i)})}{\gamma_{[\tzero,t_{n-1}]}(\tokornottok{n-1}^{(i)},\simplifiedtheta{n-1}^{(i)},y([\tzero,t_{n-1}]))\gamma_{{(t_{n-1},t_n]}}(\btokornottok{n}^{(i)},\bsimplifiedtheta{n}^{(i)}, y((t^\ast_{n-1},t_n])|\mbox{J}_s^{(i)}|}.
\end{equation*}

The particular parameter transformation $s_{1}$ should be chosen such
that if $\theta_{k_{n-1}}^{}$ and $\ttheta_{0}$ are samples from the
corresponding conditional posterior distributions, then
$\theta^{*}_n=s_{1}(\theta_{k_{n-1}}^{},\ttheta_{0})$ should approximate
a draw from the posterior for the joined segment.
The transformation $s_{2}$ is less critical, but should have a
distribution that can be loosely identified so as to
guide how to extend the target distribution. An example is provided in Section
\ref{sec:sncp}.

\subsection{Illustrative examples}\label{sec:smc.examples}

Two examples are now presented where $y(t)$ is a Poisson process. In
the first example the proposed changepoint SMC algorithm (referred to
as \US) is demonstrated on the coal-mining disaster data analyzed by
\cite{raferty86}, \cite{green05} and \cite{ajay06} among
others. Performance is compared with the SMC samplers algorithm used
in \cite{ajay06}.
The second example demonstrates the non-conjugate extension of the
algorithm using data simulated from a shot noise cox process taken from \cite{whiteley11}; performance is compared with the piecewise
deterministic processes(PDP) particle filter of
\cite{whiteley11} when applied to the same model.



For both examples, standard RJMCMC is used as an alternative method to
sample from the sequence of posteriors
$\pi_{[\tzero,t_n]}(\tokornottok{n}|y([0,t_n]))$, with the
\textit{maximum a posteriori} sample obtained from
$\pi_{[\tzero,t_{n-1}]}$ used as a starting value for sampling from
$\pi_{[\tzero,t_n]}$. For gauging performance, this ``sequential
MCMC'' (denoted SMCMC) approach can be considered as a \textit{gold
  standard} way to generate samples from the posterior given the data
from $[\tzero,t_n]$, although carrying a much higher computational
cost than SMC.




\subsubsection{Coal data}\label{sec:coal_data} 
The coal mining disaster data consist of the times of coal-mining
disasters in the UK between 1851 and 1962 and are a popular data set
for applying changepoint analysis. In a sequential time frame these
data were analyzed in \cite{ajay06}, and a comparison will be made
with results from the SMC samplers ($\AJ$) algorithm in that article
using the code provided for this example therein. It is assumed that
the disasters follow an inhomogeneous Poisson process with piecewise
constant intensity, and the piecewise constant intensity is estimated
sequentially.

Define the parameter vector $\simplifiedlambda{n}=(\lambda_{0},\ldots,\lambda_{k_n})$, such that $\lambda(t)=\sum_{i=0}^{k_n}\lambda_i\I_{(\tau_i,\tau_{i+1}]}(t)$, where $\lambda_{i}\in \mathbb{R}^{+}$ is the intensity of the process between $\tau_{i}$ and $\tau_{i+1}$. The likelihood of the observed process data is
\begin{equation*}
f(y([\tzero,t_n])|\tokornottok{n},\simplifiedlambda{n})=\prod_{i=0}^{k_n}\lambda_{i}^{r_{i}}\exp\{-\lambda_{i}(\tau_{i+1}-\tau_{i})\},
\end{equation*}
where $r_{i}=\int_{\tau_i}^{\tau_{i+1}}\diff y(t)$ is the number of events between $\tau_{i}$ and $\tau_{i+1}$.

The $(k_n+1)$ intensity levels  $\lambda_{i}$ are initially assumed to follow independent conjugate $\Gamma(\alpha, \beta)$ priors,
\begin{equation}
p(\simplifiedlambda{n}|\tokornottok{n})=
\prod_{i=0}^{k_n}\frac{\beta^\alpha}{\Gamma(\alpha)}\lambda_i^{\alpha-1}\exp(-\beta \lambda_i)\label{eq:intensity_conjugate_prior}.
\end{equation}
For posterior inference the intensities $\simplifiedlambda{n}$ can then be integrated out to obtain the posterior distribution for the changepoints \eqref{eq:posterior}, which is known only up to proportionality through
\begin{equation}
\gamma_{[\tzero,t_n]}(\tokornottok{n},y([\tzero,t_n])= \nu^{k_n}\exp(-\nu t_n)\prod_{i=0}^{k_n}\frac{\beta^\alpha}{\Gamma\left(\alpha\right)}\frac{\Gamma\left(\alpha+r_i\right)\hspace{2ex}}{\left(\beta+\tau_{i+1}-\tau_{i}\right)^{\alpha+r_{i}}},\label{eq:dist.gamma}
\end{equation}
since the normalizing constant does not have an analytic solution. Conditional on the changepoints
$\tokornottok{n}$, the intensity levels $\lambda_i$ have known independent
posterior distributions
\begin{equation}
\lambda_i | \tokornottok{n},y([\tzero,t_n]) \sim \Gamma\left(\alpha+r_i,\beta+\tau_{i+1}-\tau_i\right).
\label{eq:posteriorlambda}
\end{equation}
\cite{ajay06} also assumes the changepoints are a Poisson
process, but has a non-conjugate prior $p_{\AJ}(\simplifiedlambda{n}|\tokornottok{n})$ for the intensities by assuming
$\lambda_{0}\sim\Gamma(\alpha_{\AJ},\beta_{\AJ})$ and $\lambda_{i}|
\lambda_{i-1}\sim
\Gamma({\lambda_{i-1}^2}/{\chi},{\lambda_{i-1}}/{\chi})$.

To make directly comparable inference under the proposed SMC
algorithm, whilst still adopting the conjugate priors for the
intensity levels \eqref{eq:intensity_conjugate_prior} for ease of
sampling, the following particle re-weighting is proposed: Given a
weighted sample of changepoints
$\{\tokornottok{n}^{(i)},w_{n}^{(i)}\}_{i=1}^{N}$ obtained from
Algorithm \ref{PFSMC}, $\simplifiedlambda{n}^{(i)}$ can first be sampled  for each changepoint segment
directly from \eqref{eq:posteriorlambda}, to
give a weighted sample of changepoints and intensities
$\{\tokornottok{n}^{(i)},\simplifiedlambda{n}^{(i)},w_{n}^{(i)}\}_{i=1}^{N}$. Second,
this augmented sample can then be simply reweighted to give an approximate
sample from the non-conjugate model of \cite{ajay06}, with new
weights $\bar{w}_{n}^{(i)}$ given by
\begin{equation}
\bar{w}_{n}^{(i)}=w_{n}^{(i)}\frac{p_{\AJ}(\simplifiedlambda{n}^{(i)}|\tokornottok{n})}{p(\simplifiedlambda{n}^{(i)}|\tokornottok{n}^{(i)})}.\label{eq:reweighting}
\end{equation}
Note that for the sequential MCMC method (SMCMC) to be used as a benchmark for comparison, after first performing efficient RJMCMC on the conjugate model, the same sampling of intensities and reweighting strategy is used.

\cite{ajay06} chose to perform inference annually, which implies a sequence of $112$ changepoint densities where the $n$th density concerns the date range $[1851,1851+n]$. The chosen prior parameter for the number of changepoints $\nu={2}/{112}$; for the priors for the non-conjugate model, following \cite{ajay06} $\alpha_{\AJ}=4.5$ and $\beta_{\AJ}=1.5$, and $\chi=5$; for the conjugate model uninformative priors are chosen, $\alpha=\beta=0.1$. Again following \cite{ajay06}, the overall number of particles $N=10{,}000$. For SMCMC, $1{,}000{,}000$ samples are drawn from each posterior to give reliable posterior estimates.


\begin{figure}[t]
\centering
\includegraphics[width=1\textwidth]{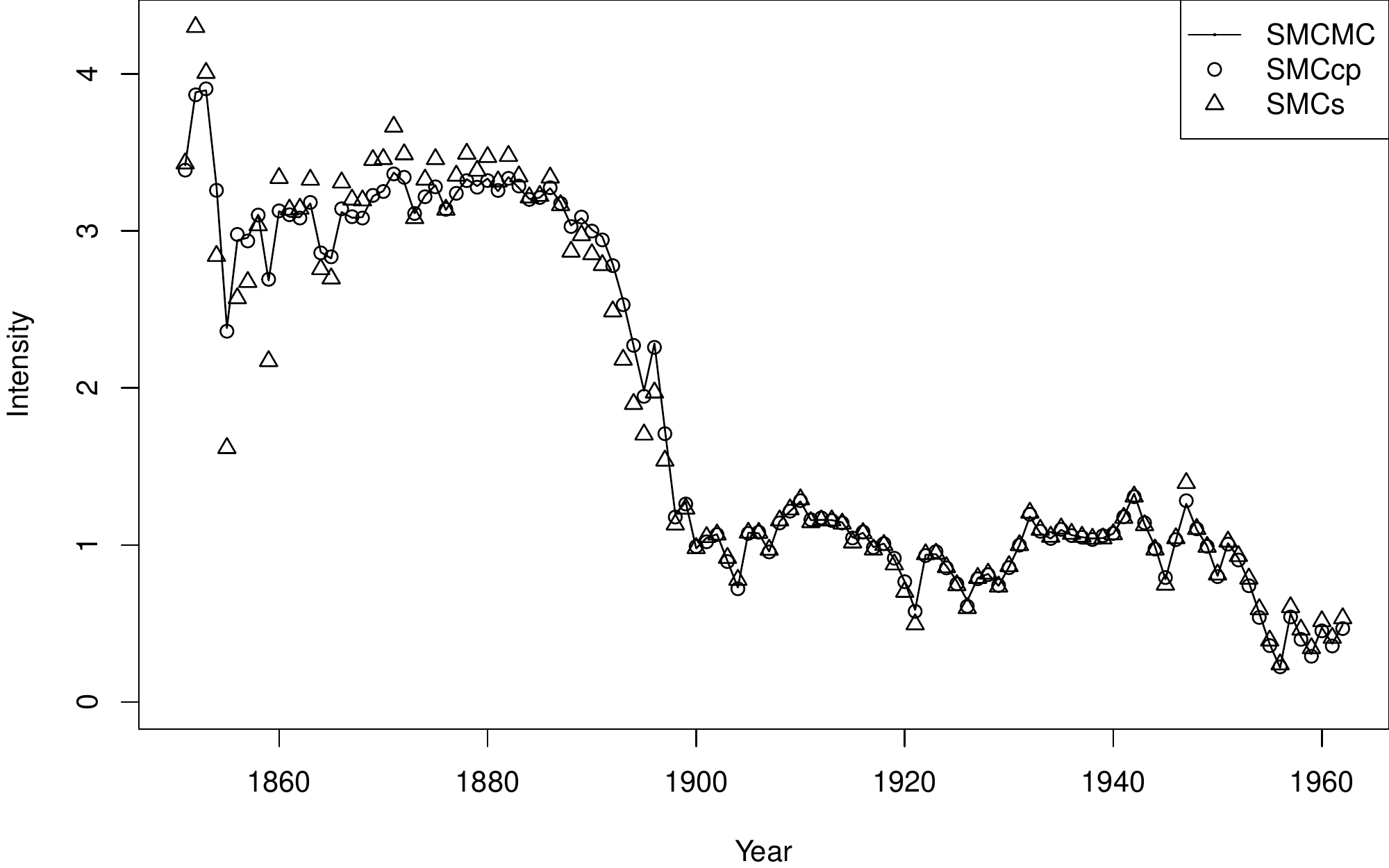}
\caption{Online estimated intensity function for the coal-mining disaster data}
\label{fig:coal_intensity}
\end{figure}

\begin{figure}[t]
\centering
\subfloat{\includegraphics[width=.49\textwidth]{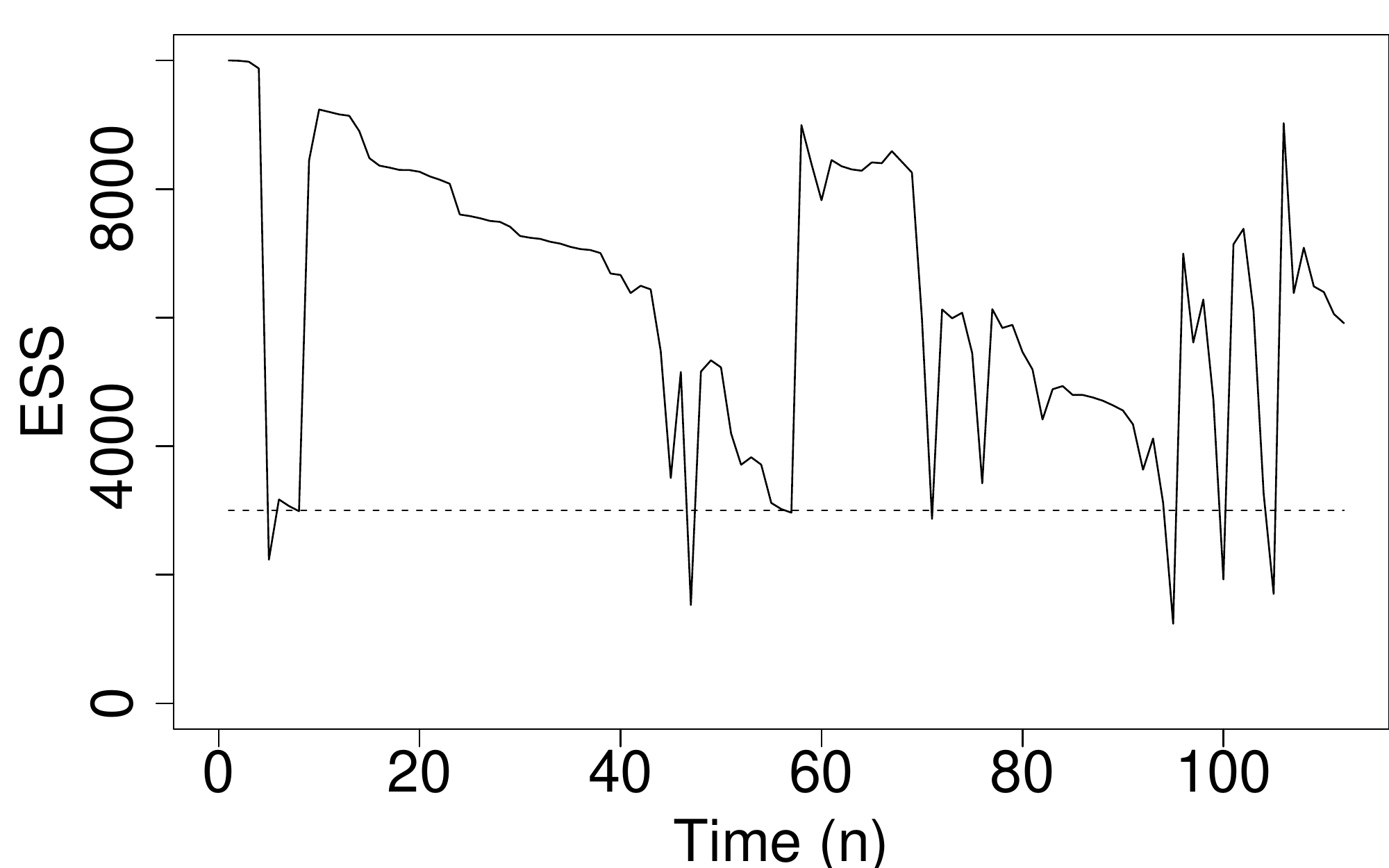}}\hspace{0.01\textwidth}
\subfloat{\includegraphics[width=.49\textwidth]{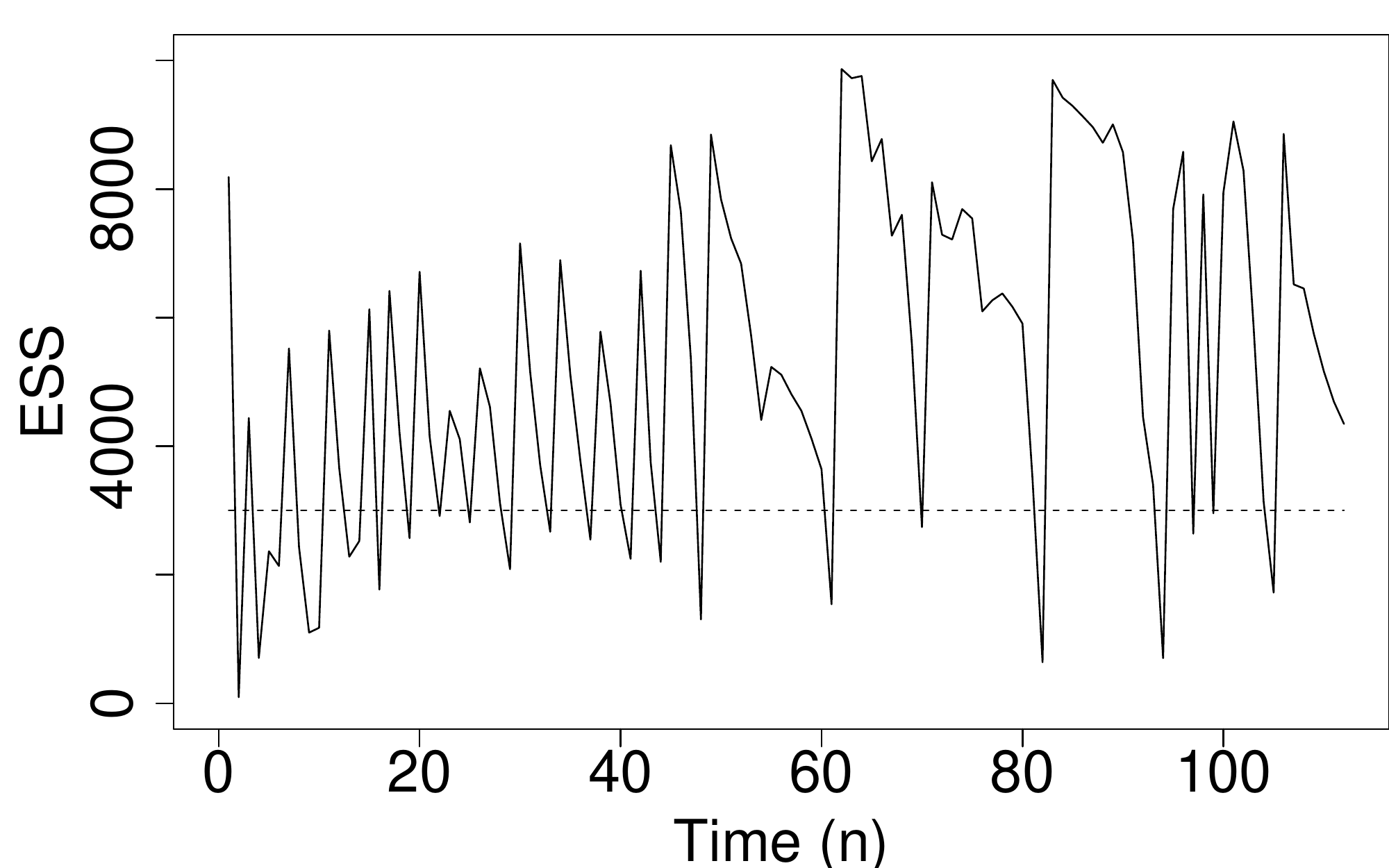}}\\
\caption[Effective sample size]{Effective sample size. Left: \US Right: \DM}
\label{fig:coal_ess}
\end{figure}

Figure \ref{fig:coal_intensity} shows the online or filtered intensity
function for the non-conjugate model, estimated each year using \US,
\DM and SMCMC. The \US method perfectly tracks the target SMCMC curve,
which represents the best possible inference. In particular, \US
performs much better than \DM for the first fifty years of densities,
after that from $1900$ the performance of both algorithms are more
comparable. Figure \ref{fig:coal_ess} shows the ESS
for both algorithms, for \US the ESS shows good stability across the $112$ year period, whereas \DM initially suffers from
persistent weight degeneracy. In total \DM performs resampling due to
the ESS dropping below the threshold $26$ times, where $19$ of the
times are before $1900$, compared with \US performing resampling only
$8$ times in total. This could partly be due to the initialization of
the particles from the prior in \DM and the high uncertainty in the
target distributions in the first few years.

\subsubsection{Shot noise Cox process}\label{sec:sncp}
Now suppose $y(t)$ is a shot noise Cox process, where changepoints
$\tau_{1:k_n}$ now correspond to shots (positive jumps) in the
intensity function
\begin{equation*}
\lambda(t)=\sum_{i=0}^{k_n}\lambda_i \exp\{-\kappa(t-\tau_i)\} \I_{(\tau_i,\tau_{i+1}]}(t),
\end{equation*}
where $\kappa>0$ is a fixed decay parameter for the decrease in
intensity between shots. The parameters $\simplifiedlambda{n}=\lambda_{0:k_n}$ are the
random intensity levels immediately after the shots, $\lambda_i\triangleq \lambda(\tau_i)$ and are
constrained such that the shots are always positive;
that is, $\lambda_i>\lambda_i^-$ where
$\lambda_i^-\triangleq\lambda_{i-1}\exp\{-\kappa(\tau_i-\tau_{i-1})\}$ is the intensity just before $\tau_i$.

Following \cite{whiteley11}, the prior density for the intensity levels is
\begin{equation}
p(\simplifiedlambda{n}|\tokornottok{n})=\alpha^{k_n+1}\exp(-\alpha \lambda_0)
\prod_{i=1}^{k_n}\exp\{-\alpha(\lambda_i-\lambda_i^-)\} \I_{(\lambda_i^-,\infty)}(\lambda_i)
\label{eq:sncp_intensity_prior}.
\end{equation}

The likelihood of the observed process data is
\begin{equation*}
f(y([\tzero,t_n])|\tokornottok{n},\simplifiedlambda{n})
=\exp\left\{-\sum_{i=0}^{k_n}(\lambda_i -\lambda_{i+1}^-)/\kappa+\int_{t=0}^{t_n}\log\lambda(t)\diff y(t)\right\}.
\end{equation*}

A conjugate model approximation to the shot noise Cox process would
need to forgo the constraint $\lambda_i>\lambda_i^-$. Particularly for
low values of $\kappa$, which constitute the harder inference
problems, reweighting samples from the unconstrained conjugate model
in SMC would be an unreliable approach, since the proportion of
particles obeying the required constraint, and therefore having
non-zero weight according to the target model, would diminish over
time. So for this application the non-conjugate algorithm of Section
\ref{sec:non-conjugate} is favored.

When performing a ``birth'' move in RJMCMC, the location of the new
changepoint is drawn from the proposal distribution specified in
\cite{whiteley11}. This simply puts higher probability of proposing a changepoint at regions in
the process $y(\cdot)$ where an increased rate of occurrence of events
was observed, which might correspond to a shot. To follow the
non-conjugate SMC algorithm, the intensity parameters $\lambda_i$ are
also sampled during RJMCMC. Note that, conditional on all other
parameters, the $\lambda_i$ corresponding to interior changepoints
($0<i<n$) have both lower and an upper constraints: Necessarily,
$\lambda_i\in(\lambda_i^-,\exp\{\kappa(\tau_{i+1}-\tau_i)\}\lambda_{i+1})$. Under
\eqref{eq:sncp_intensity_prior}, the full conditional distribution of
$\lambda_i$ is a truncated gamma distribution
\begin{equation*}
\pi(\lambda_i|\cdot)\propto \lambda_i^{r_i}\exp\{-(\alpha\kappa+1)z_i \lambda_i\}\I_{(\lambda_i^-,\exp\{\kappa(\tau_{i+1}-\tau_i)\lambda_{i+1})\}}(\lambda_i),
\end{equation*}
where $r_i$ is the number of $y$ events in $(\tau_i,\tau_{i+1}]$ and
  $z_i=[1-\exp\{-\kappa(\tau_i-\tau_{i-1})\}]/\kappa$.

For the SMC algorithm, it can be supposed that two sets of particles have been obtained at time $t_{n}$:
\begin{itemize}
\item $\{\tokornottok{n-1}^{(i)},\simplifiedlambda{n-1}^{(i)},w_{n-1}^{(i)}\}_{i=1}^{N}$, a weighted sample from $\pi_{[\tzero,t_{n-1}]}(\tokornottok{n-1},\simplifiedlambda{n-1}|y([0,t_{n-1}]))$;
\item $\{\btokornottok{n}^{(i)},\bsimplifiedlambda{n}^{(i)}\}_{i=1}^{N}$, sampled from $\pi_{(t_{n-1},t_{n}]}(\btokornottok{n},\bsimplifiedlambda{n}|y((t^\ast_{n-1},t_n]))$. 
\end{itemize}
To combine the two intensity parameters $\lambda_{k_{n-1}}$ and  $\tilde{\lambda}_{0}$ to a single intensity parameter $\lambda^\ast_n$ for the merged interval $(\tau_{k_{n-1}},\ttau_{1}]$, an attractive function would be
\begin{equation*}
s_1(\lambda_{k_{n-1}},\tilde{\lambda}_{0})=\frac{[\lambda_{k_{n-1}}(1-\exp\{-\kappa(t_{n-1}-\tau_{k_{n-1}})\})+\tilde{\lambda}_{0}(1-\exp\{-\kappa(\ttau_1-t_{n-1})\})]}{(1-\exp\{-\kappa(\ttau_1-\tau_{k_{n-1}})\})},
\end{equation*}
which preserves the cumulative intensity over
$(\tau_{k_{n-1}},\ttau_{1}]$. However, this function might propose an
  illegal intensity according to the constraints of the model. To
  ensure a legitimate proposal, it is easiest to work with the
  parameterization provided by the shots,
  $\theta_i=\lambda_i-\lambda_i^-$. Working in this parameter space,
  to guarantee a legitimate move the simplest choice of $s$ is then
  the bivariate identity function, implying
  $\theta_n^\ast=\theta_{k_{n-1}}=\lambda_{k_{n-1}}-\lambda_{k_{n-1}}^-$ and $u_{n-1}=\ttheta_0=\tilde{\lambda}_{0}$. This
  proposal has the potential to work well, since the proposal density
  for changepoints in $(t_{n-1},t_n]$ assumes the last shot in
    $[0,t_{n-1}]$ was at $t_{n-1}^\ast$, and carries forward the data
    from $t>t_{n-1}^\ast$.

 Finally, to extend the target distribution,
 $\tpi(u_{n-1}|\tau_{1:k_{n}},\lambda_{0:k_{n}})$ can be defined to be
 the full conditional from which the parameter $\ttheta_0$ was
 originally proposed.

Viewed from the intensity parameterization, the proposed parameters for the whole interval will be:
\begin{align*}
&\{\tokornottok{n}^{(i)}=(\tau_{1}^{(i)},\ldots,\tau_{\ki_{n-1}}^{(i)},\ttau_{1}^{(i)},\ldots,\ttau_{\kki_{n}}^{(i)})\}_{i=1}^{N}\\
&\{\simplifiedlambda{n}^{(i)}=(\lambda_{0}^{(i)},\ldots,\lambda_{\ki_{n-1}}^{(i)},\tlambda_{1}^{(i)}+\delta_n^{(i)},\ldots,\tlambda_{\kki_{n}}^{(i)}+\delta_n^{(i)}\}_{i=1}^{N},
\end{align*}
where  $\delta_n^{(i)}=\lambda_{\ki_{n-1}}^{(i)}\exp\{-\kappa(\ttau_1^{(i)}-\tau_{k_{n-1}}^{(i)})\}-\tlambda_0^{(i)}\exp\{-\kappa(\ttau_1^{(i)}-t_{n-1})\}$.

For a true comparison against the piecewise deterministic process
(\NW) algorithm in \cite{whiteley11}, the same shot noise parameter
values are used: $\nu={1}/{40}$, $\kappa=1/100$ and
$\alpha={2}/{3}$. The data obtained from the code provided in
\cite{whiteley11} are simulated over $[0,2000]$, with 40 update
intervals each of length 50. The total number of particles $N=500$ and
the ESS resampling threshold is set to $200$. Again as a comparison to
both SMC algorithms, the slower but accurate SMCMC algorithm is used
to provide a ``gold standard'' of inference.

\begin{figure}[t]
\centering
\includegraphics[width=1\textwidth]{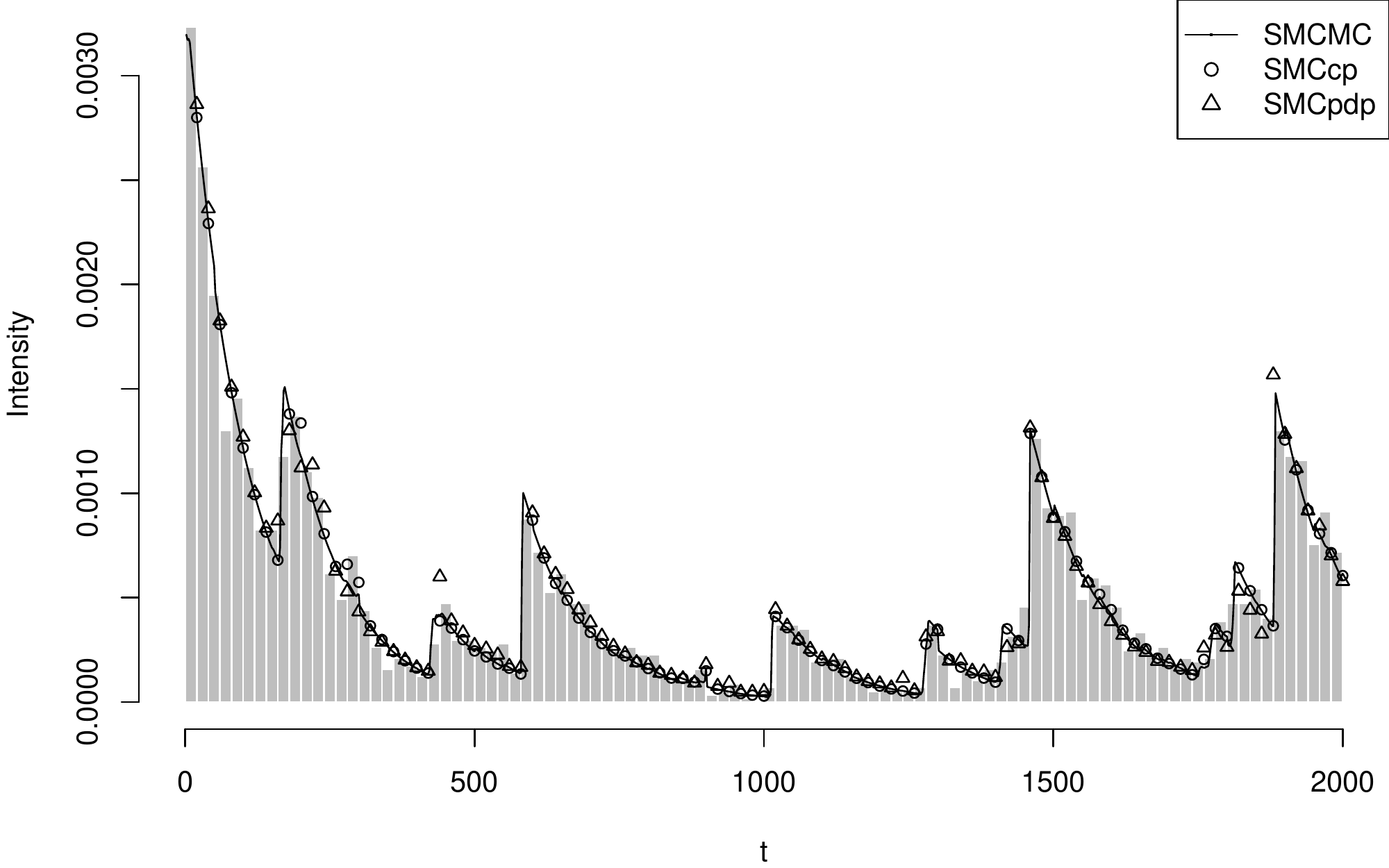}
\caption{Online estimated intensity function}
\label{fig:sncp_intensity}
\end{figure}

Figure \ref{fig:sncp_intensity} shows the online
filtering estimate for the intensity function using the three
different algorithms \US, \NW and SMCMC for the shot noise Cox process
as well as a histogram of the data over the time period. Although both SMC algorithms perform well, the
\US algorithm tracks the SMCMC curve more reliably, with \NW
slightly overestimating some of the shots in the intensity.

\begin{figure}[t]
\centering
\subfloat{\includegraphics[width=.49\textwidth]{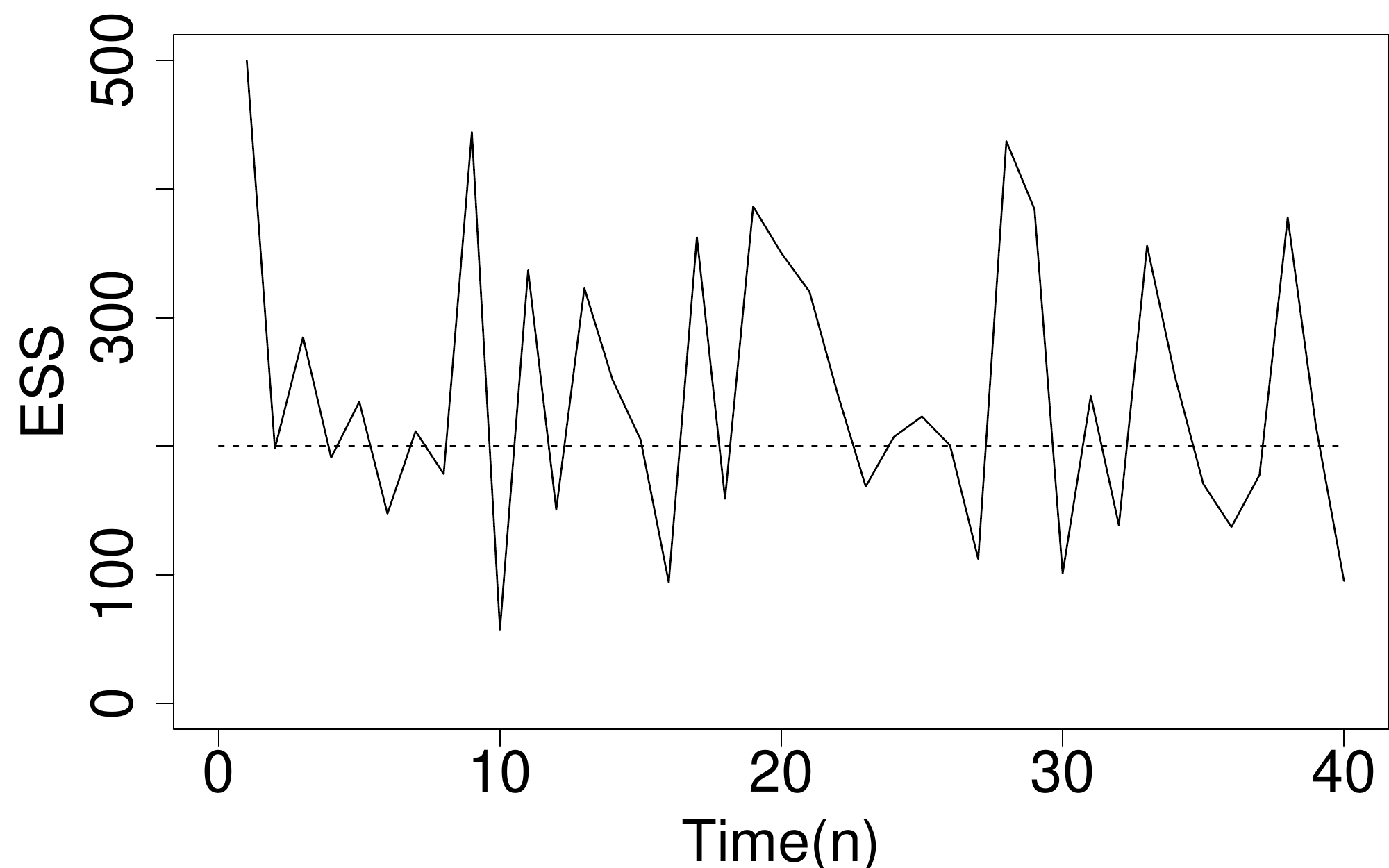}}\hspace{0.01\textwidth}
\subfloat{\includegraphics[width=.49\textwidth]{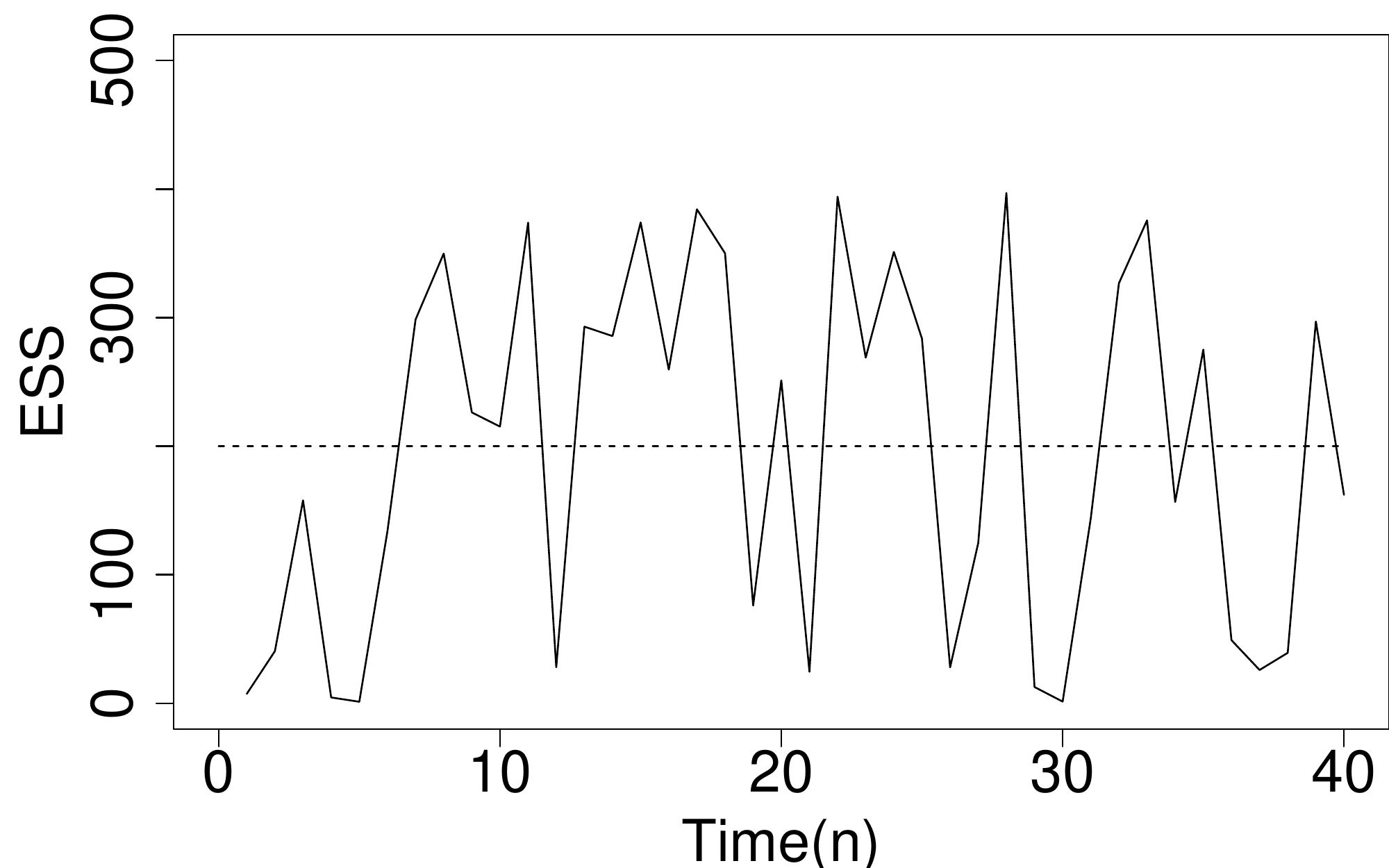}}\\
\caption[Effective sample size]{Effective sample size. Left: \US Right: \NW}
\label{fig:sncp_ess}
\end{figure}

Figure \ref{fig:sncp_ess} shows the ESS at each time point for both
\US and \NW, and \US shows much better stability in terms of the
variability of the weights. The proposal distribution specified in
\cite{whiteley11} only allows birth of changepoints within each update
interval, so initially the algorithm may be overfitting changepoints
causing high variability in the weights.

\begin{figure}[t]
\centering
\subfloat{\includegraphics[width=.49\textwidth]{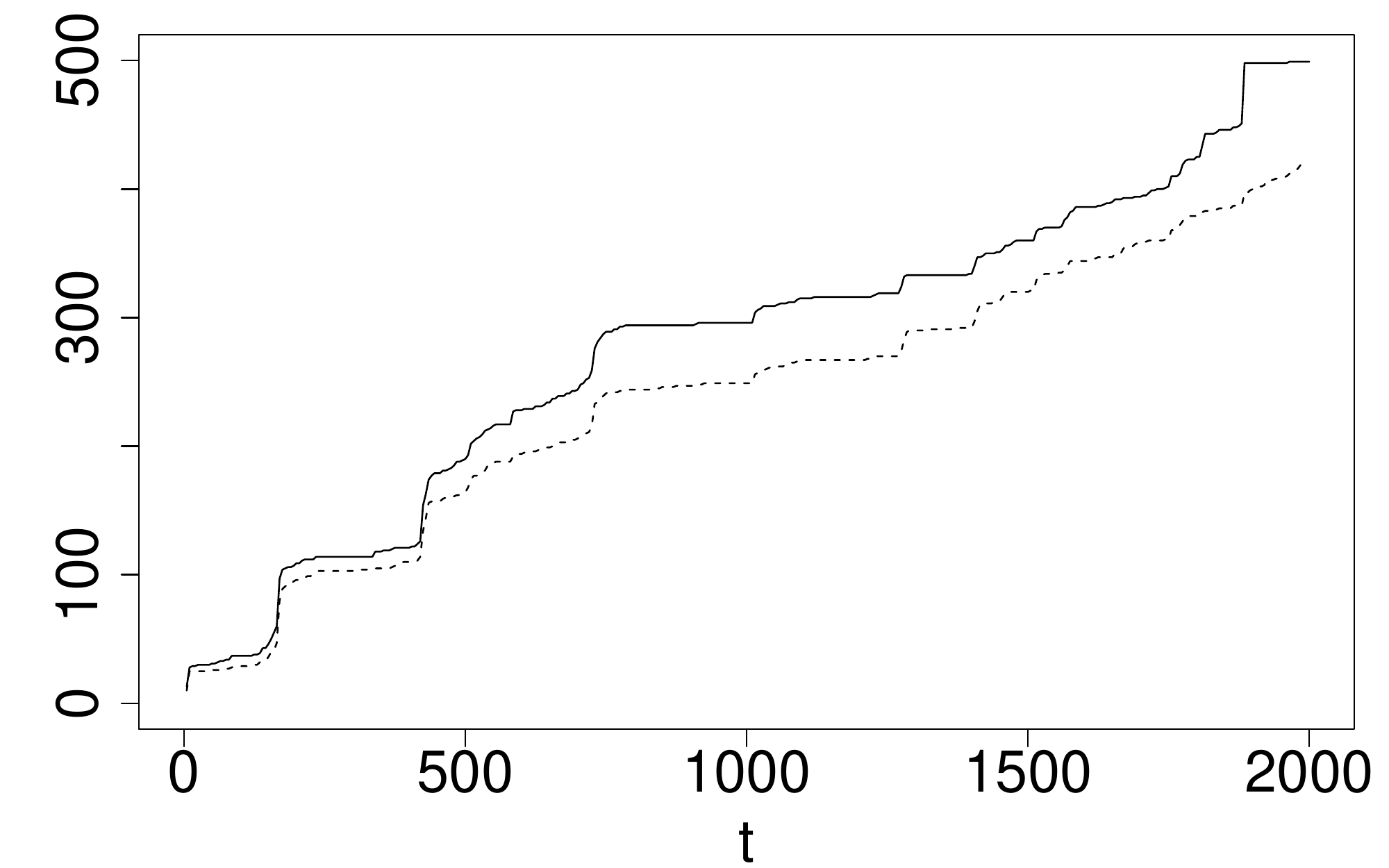}}\hspace{0.01\textwidth}
\subfloat{\includegraphics[width=.49\textwidth]{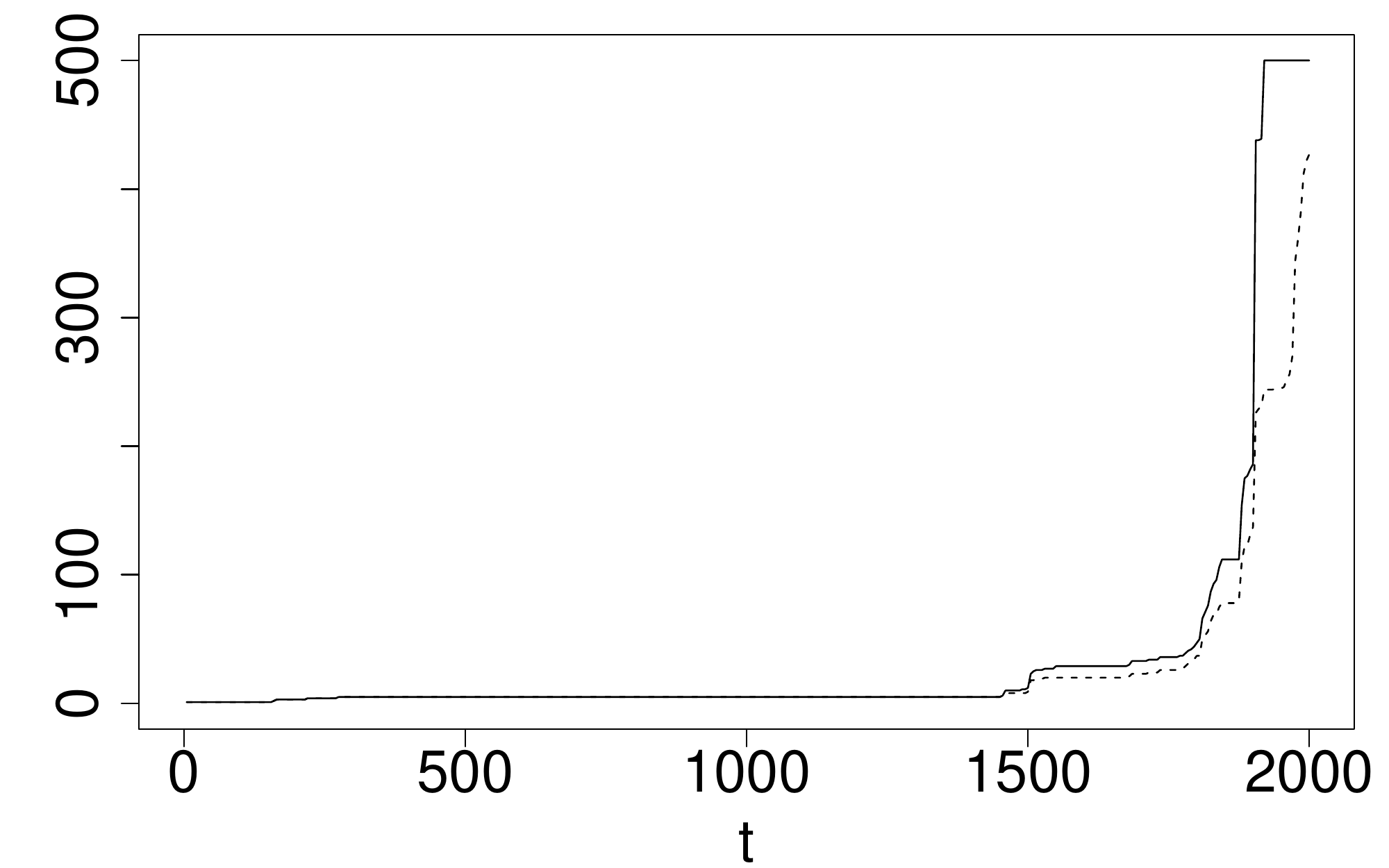}}\\
\caption[Effective sample size]{Unique particles at the final iteration versus time. Left: \US Right: \NW. Solid:pre-resampling. Dotted:post-resampling}
\label{fig:sncp_unique}
\end{figure}

\cite{whiteley11} also plotted the number of unique particles, over
time, that eventually survived to the final iteration of the SMC
algorithm. Figure \ref{fig:sncp_unique} plots this quantity both
before and after resampling, as although there may be many unique
particles the importance weights may have high variance, implying a
low quality particle approximation, \cite{whiteley11}. \US shows much
more of a diverse particle set further back in time then \NW both
before and after resampling.

\section{Adaptive sequential Monte Carlo}\label{sec:adaptive_smc}
In many applications such as finance or security, there can be cause
to make sequential inference about many independent target probability
distributions in parallel. In finance, such problems could arise in
automated trading, where beliefs about the future prices of many
stocks will be continually updated; and in security, statistical
models can be used for monitoring each entity in a large network for
unusual behavior. Notationally, suppose there is a collection of $m$ sequences
of target distributions $\{\pi^j_{[\tzero,t_n]}\}$, $j=1,\ldots,m$,
such that inference is to be made about each target distribution
$\pi^j_{[\tzero,t_n]}$ at the same sequence of update times
$t_1<t_2<\ldots$.

For SMC algorithms, when updating beliefs
about each of these target distributions 
it is desirable to allocate more computational resource to those
target distributions that appear to have changed the most. For
existing SMC methods this idea is problematic and has not been
explored, since the number of particles $N$ is fixed from the start;
these particles are either refined or resampled as the target
distribution evolves, but the same number of particles is always
maintained. Indeed, the effective sample size typically drops at
precisely those times when the target distribution undergoes the most
change, and resampling is required to ensure that the $N$ particles
are still relevant, at the cost of a loss of diversity of the
particles.

\if1\blind{\cite{heard14} derived a sequential
approach to determining sample sizes during sampling, based on the
apparent relative entropy of the different distributions. }\fi
\if0\blind{As a prequel to this article, \cite{heard14} derived a sequential
approach to determining sample sizes during sampling, based on the
apparent relative entropy of the different distributions.}\fi 
By estimating the Kullback-Leibler divergence of each sample from its
target, samples can be allocated to the target distribution with the
highest discrepancy. In this work, the adaptive sample size
strategy is applied at each update when sampling from the proposal
distributions
$\pi^j_{(t_{n-1},t_{n}]}(\btokornottok{n}|y((t^{\ast j}_{n-1},t_n]))$.

The rationale behind this approach is as follows. By conditioning on
all data since the estimated most recent changepoint $t^{\ast
  j}_{n-1}$, the proposal density
$\pi^j_{(t_{n-1},t_{n}]}(\btokornottok{n}|y((t^{\ast
    j}_{n-1},t_n]))$ was chosen such that
\begin{equation*}
\pi^j_{[\tzero,t_n]}(\cdot,\cdot|y([0,t_n])) \approx \pi^j_{[\tzero,t_{n-1}]}(\cdot,\cdot|y([0,t_{n-1}])) \pi^j_{(t_{n-1},t_{n}]}(\cdot,\cdot|y((t^{\ast j}_{n-1},t_n]))
\end{equation*}
Hence this proposal density will have higher entropy when more
probability is assigned to the existence of multiple new changepoints
of uncertain location within $(t_{n-1},t_{n}]$, which in turn implies
  a larger distance between the old and new target distributions. So
  by taking more samples during the current update interval, the
  uncertainty surrounding the new changepoints will be better
  captured. Whereas if another target distribution $j'$ strongly
  appears to have no new changepoints during the same update window,
  it will be acceptable to take fewer samples to represent this
  portion of the distribution.

  At time $t_{n-1}$, suppose there were $N^j$ weighted samples
  approximating the $j$th target distribution
  $\pi^j_{[\tzero,t_{n-1}]}$. Then, following the algorithm of
  \cite{heard14} or some other adaptive strategy, suppose $M^j$
  samples are obtained from
  $\pi^j_{(t_{n-1},t_{n}]}(\tokornottok{n}|y((t^{\ast
    j}_{n-1},t_n]))$,
  the update proposal at time $t_n$. Typically $M^j\neq N^j$, so at
  step \ref{step:combine} of the SMC Algorithm \ref{PFSMC} there are
  two unequal sized groups of particles to combine. To redress this
  imbalance, copies need to be made of some particles in the smaller
  sample so that the two sample sizes are equal. When $N^j<M^j$, this
  task of replicating particles can be used advantageously to reduce
  the variability of the weights of the old particles and increase the
  effective sample size. The next section outlines a simple procedure
  for determining how many copies to make of each particle, and how
  the particles are consequently reweighted. It can be noted that the
  same algorithm can equally be applied for increasing the number of
  new particles when $N^j>M^j$, but this trivially reduces to
  assigning $\tokornottok{n}^{(i)}=\tokornottok{n}^{(i \mod M^j)}$
  for $i>M^j$.

\subsection{Replicating particles}
Suppose there are currently $N$ particles for the region
$[\tzero,t_{n-1}]$ that need to be paired with $M>N$ particles from
$(t_{n-1},t_n]$ following an increased sampling allocation. When
  considering duplicating particles from the weighted particle set, it
  is important to note that there may already be duplicated particles,
  perhaps from previous iterations. For continuous time changepoints,
  duplicates also arise when particles with no changepoints are
  sampled.  For simplicity of notation, assume now that the weighted
  particle set has been labeled such that the first $N'<N$ particles
  are unique. For $1\leq i\leq N$, let $m_0^{(i)}$ be the number of
  replicates of $\tokornottok{n-1}^{(i)}$ in the $N$ particles, and
  define $\bar{w}^{(i)}=w_{n-1}^{(i)}m^{(i)}_0$. Then note that
  $\{\tokornottok{n-1}^{(i)},\bar{w}^{(i)}\}_{i=1}^{N'}$ is an
  equivalent representation of the full weighted particle set, since
\begin{equation*}
\sum_{i=1}^{N'}\bar{w}^{(i)}\delta_{\tokornottok{n-1}^{(i)}}(\tokornottok{n-1})\equiv \sum_{i=1}^{N}w^{(i)}_{n-1}\delta_{\tokornottok{n-1}^{(i)}}(\tokornottok{n-1}).
\end{equation*}

It is necessary to work with this reduced representation, as otherwise
the algorithm would admit the possibility of making different numbers
of copies of the same particle. Assume that each unique particle $i$
will be replicated $m^{(i)}$ times, so that
$\sum_{i=1}^{N'}m^{(i)}=M$. Then in order to minimize the sum of the
squared weights and ensure that any Monte Carlo estimates are the same
after the particle set has been increased, the revised weight for
particle $i$ is ${\bar{w}^{(i)}}/{m^{(i)}}$. The important implication
here is that replicating highly weighted particles will reduce those
weights, which will make the weights more uniform and therefore boost
the effective sample size in step \ref{step:ess} of Algorithm
\ref{PFSMC}.

Choosing optimal values $\{m^{(i)}\}_{i=1}^{N'}$, 
so that the resulting sum of squared weights is minimized is a complex
optimization problem, and solving this directly would add too much
computational burden to the overall SMC algorithm. So instead,
Algorithm \ref{alg:particles} presents a sequential optimization
method.

\begin{algorithm}[t]\caption{Increasing particle set from N to M}{\label{alg:particles}}
\begin{algorithmic}[1]

\State Set $m^{(i)}=m^{(i)}_0$ and $\bar{w}^{(i)}=m^{(i)}_0 w^{(i)}$ for $i=1,\ldots,N'$. Let $m=\sum_{i=1}^{N'}m^{(i)}=N$

\State Calculate $\delta_i={(\bar{w}^{(i)})^2}/\{(m^{(i)}+1)m^{(i)}\}$  for $i=1,\ldots,N'$\label{calculate_delta}
\State Let $\istar=\argmax_i\{\delta_i:i=1,\ldots,N'\}$
\While { $m < M$}

\State Let $\istartwo=\argmax_i \{\delta_i:i=1,\ldots,N' {,} i\neq \istar\}$
\State \begin{varwidth}[t]{13.8cm} Let
$x_{N'}=\min\left(M-m,\lceil\sqrt{{(\bar{w}^{(\istar)})^2}/{\delta_{\istartwo}}+0.25}-0.5-m^{(\istar)}\rceil\right)$\label{calculatex}
\end{varwidth}
\State  Let $m^{\istar}=m^{\istar}+x_{N'}$ and $m=m+x_{N'}$
\State  Let $\delta_{\istar}={(\bar{w}^{(\istar)})^2}/\{(m^{(\istar)}+1)(m^{(\istar)})\}$ 
\State Let $\istar=\istartwo$
\EndWhile
\State Let $i'=1$
\For {$i=1:N'$}
\For {$j=1:m^{(i)}$}
\State $\tokornottok{n-1}^{*(i')}=\tokornottok{n-1}^{(i)}$
\State $w_{n-1}^{*(i')}={\bar{w}^{(i)}}/{m^{(i)}}$
\State $i'=i'+1$
\EndFor
\EndFor
\State $(\tokornottok{n-1}^{(i)},w_{n-1}^{(i)})\leftarrow(\tokornottok{n-1}^{* (i)},w_{n-1}^{*(i)})$ for $i=1,\ldots,M$
\end{algorithmic}
\end{algorithm}

The quantity $\delta_i$, calculated in step \ref{calculate_delta},
represents the decrease in the sum of the squared weights if the $i$th
particle is replicated once, and this is used to identify the next particle to replicate, $\istar$.
The number of replicates of particle $\istar$ that are then made,
$x_{N'}$, calculated in step \ref{calculatex}, is the largest integer
solving the inequality
\begin{equation*}
\frac{(\bar{w}^{(\istar)})^2}{(m^{(\istar)}+x_{N'}+1)(m^{(\istar)}+x_{N'})}<\delta_{\istartwo},
\end{equation*}
since this is the smallest number of replicates that are required for
$\istar$ not to remain the optimal particle to replicate.

\subsection{Example: The VAST data}\label{sec:vastdata_smc}
The IEEE VAST 2008 Challenge data are synthetic data comprising information of mobile call records for a small community of $400$ mobile phones, over a $10$ day period. The challenge was aimed at social network analysis, with the aim of uncovering anomalous behavior within the social network. The data can be obtained from \url{http://www.cs.umd.edu/hcil/VASTchallenge08}.

A successful approach taken in \citep{heard10} is to monitor the incoming call patterns of each phone to detect changes from their normal patterns, and thereby obtain a much smaller subset of potentially anomalous IDs that can further be investigated. After correcting for diurnal effects on normal behavior as in \cite{heard10}, this approach reduces to the online detection of changepoints of $400$ processes which can be assumed to follow a Poisson process with a conjugate prior for the intensity as detailed in Section \ref{sec:coal_data}. Furthermore, it was later shown in \cite{heard14} that for a fixed computational effort, more accurate inference could be obtained from the Poisson process model of these data by using the adaptive sampling strategy presented in that paper.



The SMC Algorithm \ref{PFSMC} can be deployed to simulate a real time changepoint analysis of the incoming call data for each phone number in the network. Each phone number is reanalyzed each hour over the ten days of data, which corresponds to $240$ update intervals.
Furthermore, to illustrate the adaptive version of the SMC algorithm, a variable number of particles are assigned to each process in each update window according to the complexities of their distributions using the Algorithm given in \cite{heard14}.
On each interval, each process is given a minimum of 500 particles, but the total number of samples to be adaptively allocated across the processes  $\ns^{\ast}=4{,}000{,}000$; so equal sample sizes would correspond to 10{,}000 particles for each process. 

\begin{figure}[t]
\begin{center}
\includegraphics[width=1.0\textwidth]{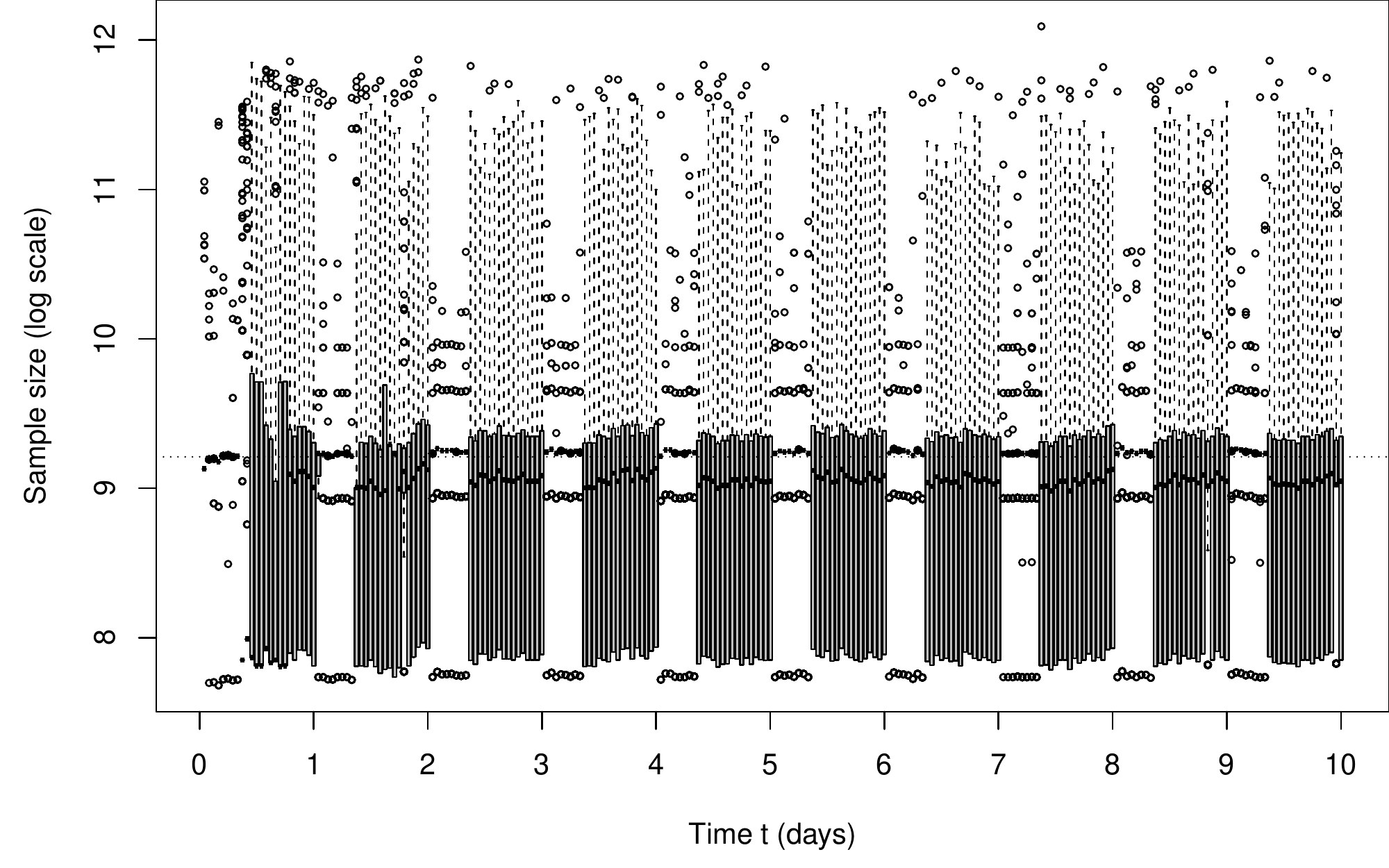}
\caption[Box plot of the sample sizes allocated to each individual on each update interval]{Box plot of the sample sizes $N^{j}$ allocated to each individual on each of the $240$ update intervals. Dotted line corresponds to $N^{j}=10{,}000$.}
\label{fig:sample_sizes_box}
\end{center}
\end{figure}

Figure \ref{fig:sample_sizes_box} shows a box plot of the sample sizes $N^{j}$ allocated to each of the $400$ processes over each update window. The dotted line shows the sample size that would be allocated to each process under a fixed sample size strategy, where $N^{j}=10{,}000$. It is apparent that at night, when most of the actors are quiet, the sample sizes allocated are similar to that of an equal sample size strategy; whereas during the day the sample sizes allocated are far more varied as actors become more active, to varying degrees. The few processes that receive larger sample sizes at night most often correspond to those actors that surprisingly become active at night, suggesting a more complicated distribution during that update interval.

\begin{figure}[t]
\centering
\subfloat{\includegraphics[width=.49\textwidth]{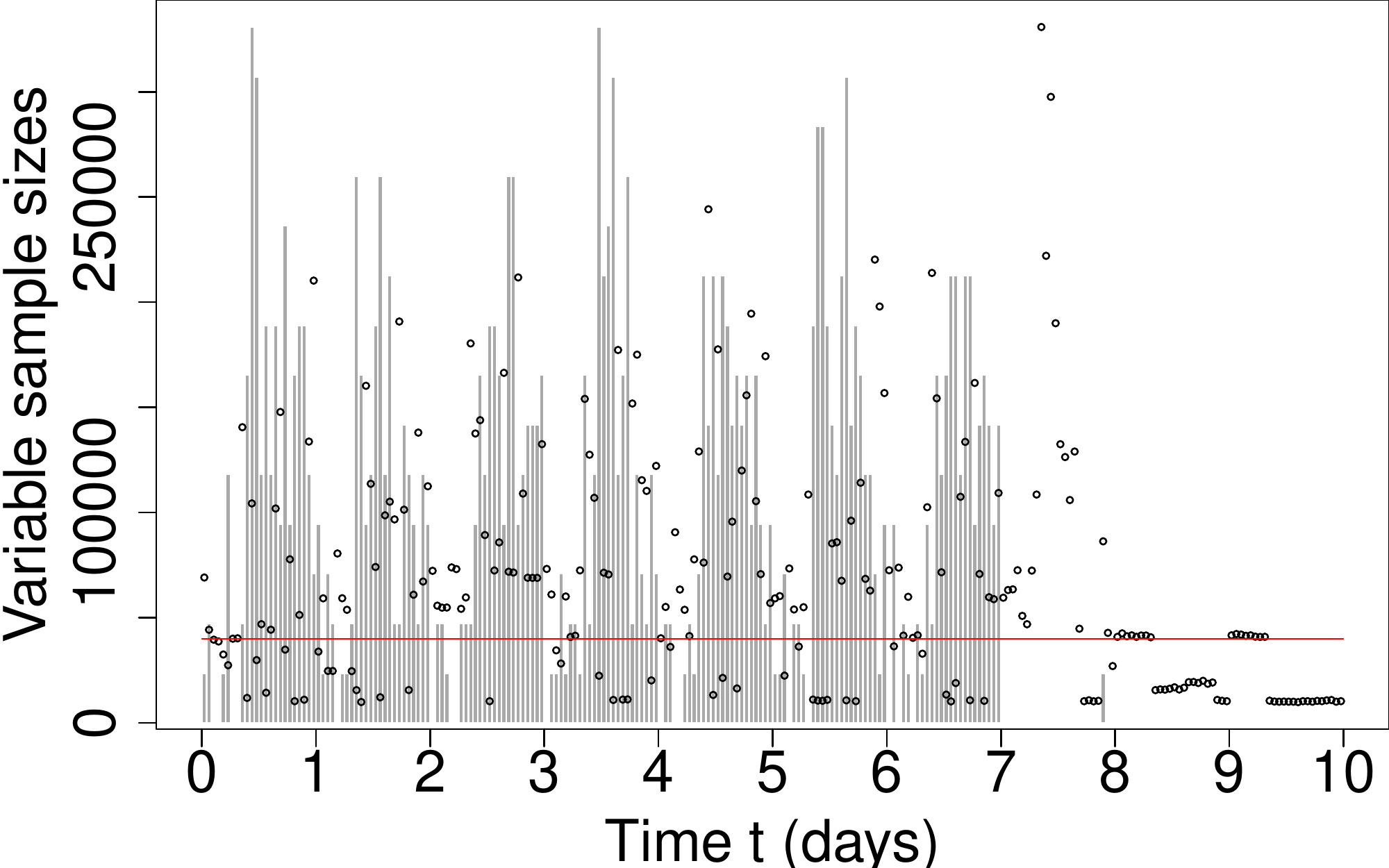}}\hspace{.01\textwidth}
\subfloat{\includegraphics[width=.49\textwidth]{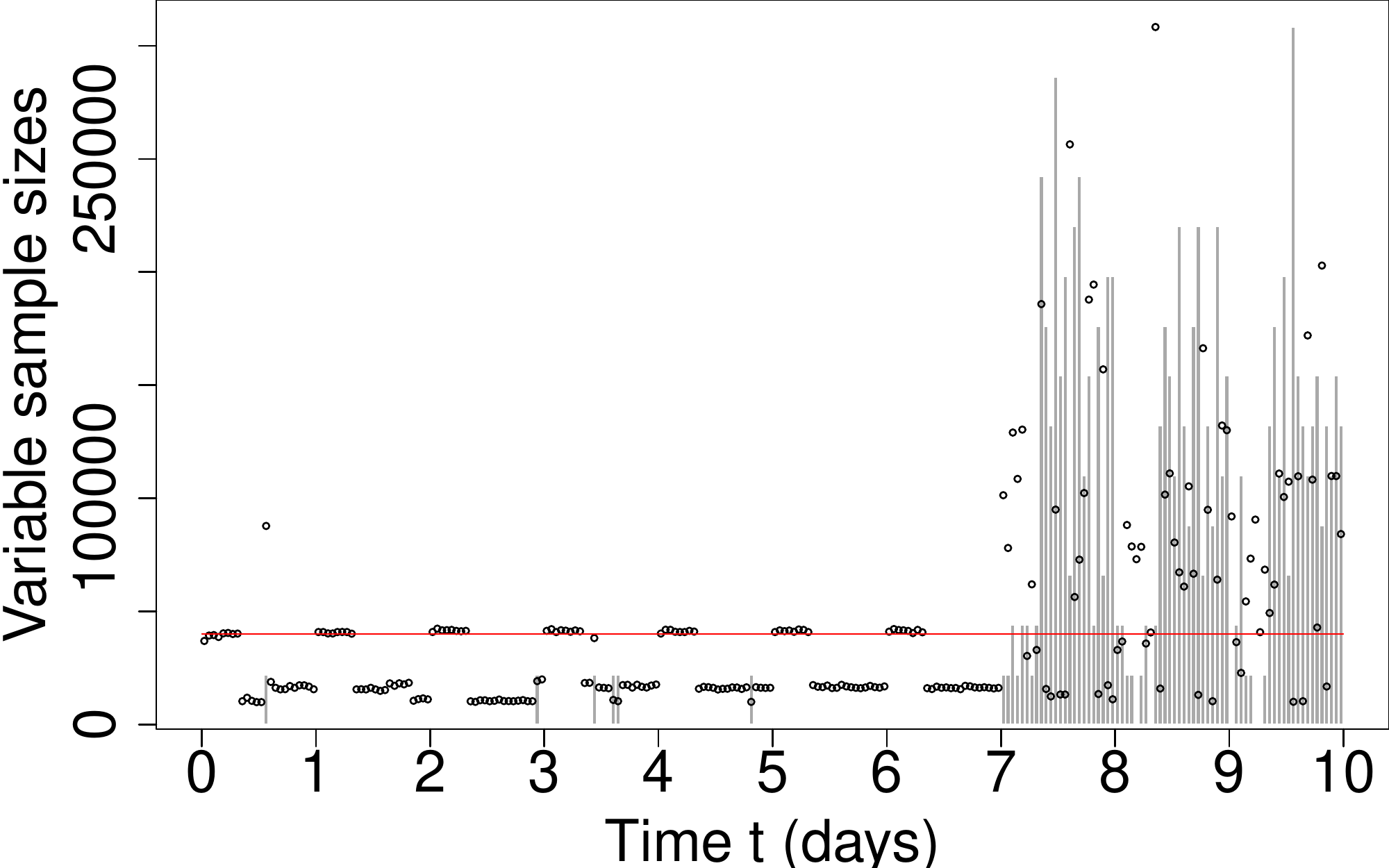}}
\caption[Allocated sample sizes for each update interval]{Total allocated sample sizes for each update interval for the processes corresponding to the unique phone IDs in Group A(left) and Group B(right). The histogram is the total number of incoming calls for the interval. The red line corresponds to the fixed sample size strategy of $N^{j}=10{,}000$.}
\label{fig:sample_sizes_bad}
\end{figure}

In \cite{ye_vast2008} it shown that the main anomalous activity was identified as involving four actors that change their handsets on the eighth day of the data period. Community members who initially communicated with phone IDs $1$, $2$, $3$ and $5$ (Group A) switched to communicating with IDs $309$, $397$, $360$ and $306$ (Group B) so that there is a decrease (and corresponding increase) in the call patterns for the handset IDs in Group A (and B). The impact of the locally anomalous behavior on sample size allocation can be seen clearly in Figure \ref{fig:sample_sizes_bad}, which shows the total allocated sample sizes for both groups over each of the update intervals. Particularly for Group A, their highest sample sizes are observed at the time of the anomaly, meaning more statistical effort is being correctly afforded to the most interesting cases.


\section{Conclusion}\label{sec:conclusion}

 A new SMC algorithm for changepoint analysis has been presented, and
 shown to outperform existing SMC methods. The computational effort of
 the algorithm does not increase over time. Effective sample size
 (ESS) thresholding has been used to control diversity of the
 particles in all examples; other standard techniques for improving
 SMC performance can also be applied to Algorithm \ref{PFSMC}, such as
 the Resample-Move algorithm of \cite{gilks01}, where MCMC transition
 kernels are applied to the particle set after ESS resampling to
 introduce diversity. The algorithm has also been shown to be adaptive
 in the number of particles used over time, which further improves
 upon the computational savings that SMC methods offer.

\bigskip
\begin{center}
{\large\bf SUPPLEMENTARY MATERIAL}
\end{center}
The C++ code and the data for the examples in Section \ref{sec:coal_data} and \ref{sec:sncp} is available to download from \url{https://github.com/mjmt05/rjmcmc.git}.

\bibliographystyle{apalike}
\bibliography{smc_paper}

\end{document}